\documentclass[prb,aps,preprint,epsfig,eqsecnum]{revtex4}
\usepackage[utf8]{inputenc}
\usepackage[T1]{fontenc}
\usepackage{graphicx}
\usepackage{longtable}
\usepackage{soul}

\def\be{\begin{equation}} 
\def\ee{\end{equation}}
\def\bea{\begin{eqnarray}} 
\def\eea{\end{eqnarray}}
\def \line{\hbox to \hsize}    
\def\frac #1#2{{#1\over #2}}

\def \ket #1{{\vert #1\rangle}}
\def \bra #1{{\langle #1\vert}}

\def\1{\mbox{\bf 1}}
\newcommand{\comment}[1]{}

\newcommand{\details}[1]{}
\def \prb {{Phys. Rev. B}}

\begin{document}
\title{Characterization of 3d topological insulators by 2d invariants}

\author{Rahul Roy \\}
\address{Rudolf Peierls Centre for Theoretical Physics, 1 Keble Road, Oxford, OX1 3NP, UK }
%\date{02 October 2009}

\begin{abstract}
  The prediction of non-trivial topological phases in Bloch insulators
  in three dimensions has recently been experimentally verified. Here,
  I provide a picture for obtaining the $Z_{2}$
  invariants for a three dimensional topological insulator by deforming
  suitable   2d planes in momentum space and by using a
  formula for the 2d $Z_{2}$ invariant based on the Chern number. The
  physical interpretation of this formula is also clarified through
  the connection between this formulation of the $Z_{2}$ invariant and
  the quantization of spin Hall conductance in two dimensions.
\end{abstract}

\maketitle 

\setcounter{tocdepth}{3}
%\tableofcontents
\vspace*{1cm}

Insulators with weak inter-electronic interactions in crystalline
materials are well described by band theory. The energy eigenstates
can be grouped into Bloch bands and the presence of a gap at the Fermi
energy prevents charge transport in the bulk since filled bands do not
contribute to electronic transport.  Since the discovery of the
quantum Hall effect, \cite{klitzing_new_1980} however, it has been
known that materials which have a bulk gap in their single particle
spectra may nevertheless have gapless edge modes, through which charge
transport may take place. Various theoretical studies, an incomplete
list of which includes
Refs. \onlinecite{laughlin1981qhc,hatsugai1993cne,thouless1982qhc,halperin_quantized_1982,PhysRevLett.61.2015},
have helped to explain the quantization of the conductance, the
connection between the bulk Hamiltonian and the Hall conductance and
the robustness and role of the edge states in the quantum Hall effect.

Recently, a set of novel topological phases which are somewhat similar
to the quantum Hall phases have been proposed to exist in ordinary
insulators with unbroken time reversal symmetry(TRS). (From here on,
unless stated otherwise, we shall restrict ourselves to a discussion
of band insulators with unbroken time reversal symmetry.) In two
dimensions, there are two distinct topological phases
\cite{kane2005zto} characterized by a single $Z_{2}$ invariant, while
in three dimensions, sixteen topological phases characterized by four
topological $Z_{2}$ invariants
\cite{roy_topological_2009,moore2006tit,fu2007tit} have been proposed
to exist in non-interacting systems. These topological phases have
been detected in HgTe quantum wells surrounded by CdTe in two
dimensions \cite{konig_quantum_2007,bernevig_quantum_2006} and in
Bi$_{1-x}$Sb$_{x}$ alloys in three dimensions \cite{fu2007tii,hsieh08}
among other compounds.  Like in the quantum Hall effect, the
topologically non-trivial phases of insulators with TRS have gapless
edge modes which are robust to small perturbations. Topological phase
transitions between these phases have been studied in
Ref. \onlinecite{murakami_universal_2008}.

While in two dimensions, the analogy with the integer quantum Hall
effect leads to a simple picture for the topological phases and the
$Z_{2}$ invariant, in three dimensions, the picture is far more
complicated. This is in part, due to the fact, that unlike the integer
quantum Hall effect, in 3d materials with unbroken TRS, while three of
the $Z_{2}$ invariants are analogous to the 2d $Z_{2}$ invariants,
there is a fourth $Z_{2}$ invariant which is intrinsically three
dimensional.

 Insulators with a non-trivial value of this fourth invariant have a
 description in terms of an effective field theory containing a
 $\theta$ term which successfully describes the physics of the
 boundary \cite{2008PhRvB..78s5424Q}. In this work, I provide a
  different argument for  counting the number of different topological
 invariants and thus the number of different topological phases in
 three dimensional insulators. Since the picture relies on
  invariants for two dimensional topological insulators, I will briefly
  review this case as well. A physical interpretation of the invariant as 
  formulated in terms of Chern numbers in two dimensions is also 
  provided.

\section{Two Dimensions}

  The original approach to the topological invariant in
two dimensions was based on counting the zeroes of a function calculated from the 
Pfaffian of a certain matrix \cite{kane2005zto}. A different approach developed by the
author was based on the obstruction of the vector bundle of
wave-functions on the torus to being a trivial bundle \cite{roy-z}.
This yielded a formula involving the Chern number of half the occupied
bands. Subsequent work using the obstructions approach and physical
ideas related to charge polarization led to a formula involving an
integral of the Berry curvature of all the bands, but restricted to
half the Brillouin zone (also called the effective Brillouin zone(EBZ))
\cite{fu2006trp}. In other related work, the topological phases in two
dimensions were studied by deforming  maps from the EBZ  to the space of 
Bloch Hamiltonians with unbroken TRS, $\mathcal{C}$, to
maps from a sphere or a torus to $\mathcal{C}$ (Ref. \onlinecite{moore2006tit}).
 The formula involving the integral 
over the EBZ in Ref. \onlinecite{fu2006trp} has been adapted to numerical 
evaluations \cite{fukui2007qsh}, while the formula in terms of the 
Chern number has the advantage of having a direct link to the edge
state physics and the integer quantum Hall effect. A brief review of 
the latter formulation is provided below.

The Hamiltonian of a Bloch insulator has single particle eigenstates,
which are either occupied or unoccupied depending on their energy
relative to the Fermi energy. The spectral projection operator is
defined as the operator that projects single particle states onto the
space of occupied states. It can thus be written as sum, $P
=\sum_{i}\ket{u_{i}}\bra{u_{i}}$, where the $\ket{u_{i}}$ are the
occupied eigenkets.  Thus the projection operator can be written as:
\(P = \sum_{k_{x},k_{y}} P(k_{x},k_{y})\) where the sum is over
reciprocal lattice vectors lying in the Brillouin zone.

 As was argued in Ref. \onlinecite{roy-z}, in materials with time reversal symmetry the
spectral projector, $P$ can be written in the form:
\begin{equation}
\label{eq:2}
 P(k_{x},k_{y}) = P_{1}(k_{x},k_{y}) + P_{2}(k_{x},k_{y})
\end{equation}
where the operators, $P_{1}, P_{2}$ are well-defined, continuous functions of the momentum
variables and are related through time reversal symmetry:
\[
 P_{1}(k_{x},k_{y})= \Theta P_{2}(-k_{x},-k_{y}) \Theta^{-1}
\]
Here, $\Theta $ is the time reversal operator which acts on the spin
degrees of freedom and is anti-linear. The choice of $P_{1}$ and
$P_{2}$ is not unique. We further assume that   $P_{1}, P_{2}$  can be chosen to be
globally smooth. (See Refs.~\onlinecite{roy-z,prodan09a} .)

For every value of $\mathbf{k}$, $P(k)$ projects onto a complex vector
space. We thus obtain a vector bundle over momentum space. The first
Chern number of this bundle, which we denote by \(\mu(P)\) can be
written in the form \cite{avron1983hqc},
\[
 \mu(P) = \frac{1}{2\pi i}\int dk_{x} dk_{y} Tr \left(  P \left(\frac{\partial
 P}{\partial k_{x}} \frac{\partial P}{\partial k_{y}} -\frac{\partial
 P}{\partial k_{y}} \frac{\partial P}{\partial k_{x}} \right) \right) .
\]

The first Chern number vanishes for $P$ in topological insulators due
to time reversal symmetry, i.e., $\mu(P)=0$. However, it was shown that
\(\nu(P)=|\mu(P_{1})|\, \textrm{mod}\, 2 \) is a topological invariant
\cite{roy-z} and is independent of the choice of $P_{1}$.

Consider the subset of gapped Bloch Hamiltonians which conserve
the $z$ component of the spin, i.e., Hamiltonians in which there are no terms which
turn up spins into down spins. The energy eigenstates can be
decomposed into spin up and spin down bands.  Hamiltonians in this
class with a bulk gap will always have a quantized spin Hall
conductance (which could be zero).

 The $Z_{2}$ topological classification tells us that Hamiltonians of
the even and the odd quantized spin Hall effects are distinct even
when we allow spin mixing terms in the Hamiltonian. In terms of the 
spin Chern number $C_{s}$ which can be defined for such models \cite{sheng2006qsh},
the $Z_{2}$ invariant is $|C_{s}/2| \textrm{mod} 2 $.

In other words, if we allow terms which cause spin mixing, the
Hamiltonians with an even spin Hall effect can all be transformed into
one another through adiabatic changes in parameter space and those
with an odd spin Hall effect can similarly be adiabatically
transformed into one another.  However, no member of the odd spin Hall
conductance class can be transformed into any member of the even spin
Hall conductance in a continuous way such that TRS is protected and
the system is gapped at all points of the transformation.  Further any
general Bloch Hamiltonian (with the same Hilbert space and which has
the same number of occupied bands) which preserves time reversal
symmetry may be adiabatically transformed to Hamiltonians of precisely
one of the two sets of Hamiltonians without breaking TRS at any
intermediate point. The trivial and non-trivial topological classes
may be thought of as equivalence classes of Hamiltonians which contain
respectively members which display an even and an odd quantized spin
Hall conductance.

\section{Three Dimensional Insulators}
\label{sec-1.6}

The Brillouin zone for a 3D insulator has the topology of a three
dimensional torus. We represent it by a cube $\{ -\pi \le
k_{x},k_{y},k_{z}\le \pi\} $. Under 
the operation of the TRS operator,
a Bloch wave-function at the point $\mathbf{k}$ gets mapped to the
point $-\mathbf{k}$. The plane in momentum space, $k_{z}=0$ 
gets mapped onto itself under inversion and has the topology of a 2d torus.
The spectral projector, $P$ for the 3d insulator restricted to this
plane, therefore has an associated $Z_{2}$ invariant.

 There are a number of such surfaces with which one may associate a
$Z_{2}$ invariant. A few of these are the planes \(k_{x}=0,
k_{x}=\pi,k_{y}=0,k_{y}=\pi,k_{z}=0 \) 
and \( k_{z}=\pi\). The associated $Z_{2}$ invariants are denoted by $\nu_{1},
\tilde{\nu}_{1}, \nu_{2}, \tilde{\nu}_{2}, \nu_{3}$ and
\(\tilde{\nu}_{3} \) respectively. It was argued previously that the $Z_{2}$ invariants of these planes
are not all independent
\cite{roy_topological_2009,moore2006tit,fu2007tit}. The arguments
were based on the counting of monopole charges
\cite{roy_topological_2009}, on contractions of the 3d EBZ to the 3d
torus \cite{moore2006tit}, or on the number of independent choices of
time reversal polarizations \cite{fu2007tit}.
Here, we provide a simple alternate argument for the number of
independent $Z_{2}$ invariants in three dimensions. This argument 
also shows how the $Z_{2}$ invariants of planes such as
$k_{x}+k_{y}=0$ may be calculated from the other $Z_{2}$ invariants.

\begin{figure}
\includegraphics[width=1\textwidth]{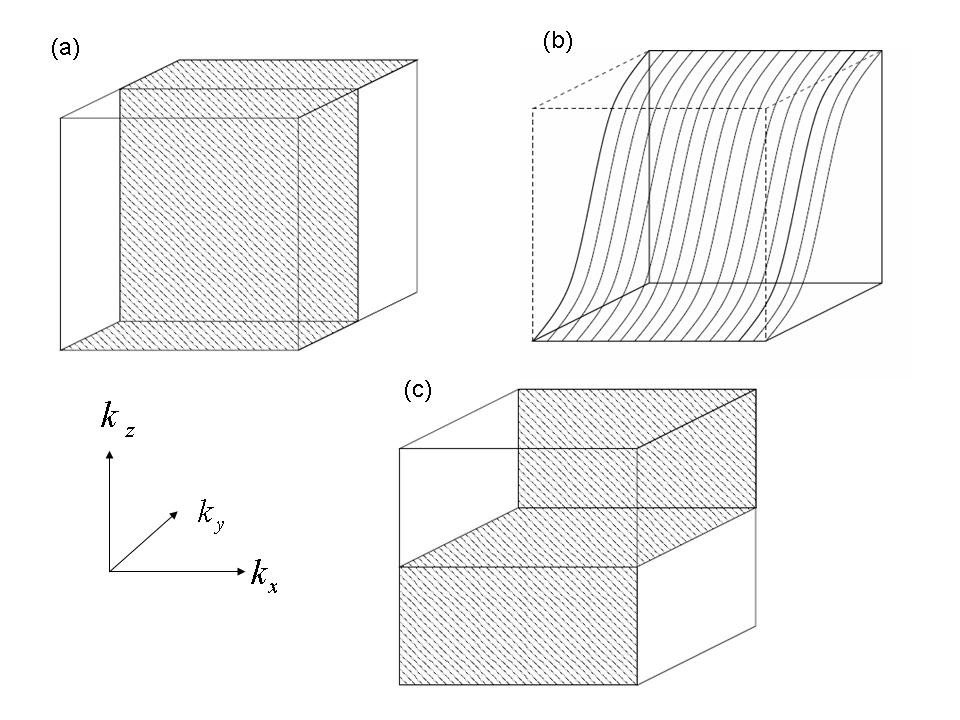}
\caption{The shaded region, S in (a) is the union of the planes $k_{y}=0$ and
$k_{z}=\pi$. This region can be continuously deformed into the shaded
regions in (b) and into the union of the planes $k_{y}=\pi$ and $k_{z}=0$ shown
in (c).}
\label{fig1}
\end{figure}

Consider the composite surface, $S$ consisting of the shaded region in
Fig.~\ref{fig1}(a). This surface, which is a union of the two planes,
$k_{y}=0$ and $k_{z}=\pi$ can be mapped onto a two dimensional torus
and is mapped onto itself under inversion.  Thus, a $Z_{2}$ invariant may be
associated with this surface. The projection operator for this surface
can be written in terms of the two projection operators as:
\[ P = P' \oplus P''
\] 
where $P'$ is the projection operator restricted to the plane, $k_z=\pi$ and  
and $P''$ is restricted to the plane $k_{y}=0$. Here, if 
$P' = \sum_{\alpha \in S'}\ket{\alpha}\bra{\alpha}, P''=\sum_{\alpha \in S''}\ket{\alpha}\bra{\alpha}$, by $P \oplus P'$, we mean $\sum_{\alpha \in S' \cup S''}\ket{\alpha}\bra{\alpha}$.

$P$, $P'$ and $P''$ can be decomposed as in Eq.~(\ref{eq:2}) and
corresponding $Z_{2}$ invariants, $\nu(P')\,,\, \nu(P'')$ defined.
The projection operator, $P$ can be decomposed as in Eq.~(\ref{eq:2})
and written as $P = P_1 + P_{2}$ where $P_1 = P'_{1} \oplus P'' _{1}$
and \( P' = P' _1 \oplus P' _2 \) etc. where we again assume that the
decomposition is such that $P_{1},P'_{1},P''_{1}$ etc. are globally
smooth on their domains of definition so that the corresponding Chern
numbers are well defined.

 It follows that 
\begin{equation}
\label{eq2}
 \nu(P) = \nu(P') + \nu(P'')  = \nu_{2} + \tilde{\nu_{3}} .
\end{equation}

Consider now, a continuous deformation of the surface
parameterized by a variable, $t$, which varies from 0 to 1 such that at
every point of the deformation, the surface $S(t)$ gets mapped onto
itself under inversion. $S(0)$ corresponds to the surface previously denoted
by S.  A possible value of $S(t)$ for \(0 < t < 1\) is shown in
Fig.~\ref{fig1}(b). At $t= 1$, the surface $S(t)$ becomes the surface
shown in Fig.~\ref{fig1}(c).  This surface is the union of the planes
$k_{y}= \pi$ and $k_{z}=0$. Under this transformation, the projection
operator which becomes a continuous function of t, $P(t)$ also changes
continuously and the $Z_{2}$ invariant therefore does not change. (Note however,
that we do not require that $P_{1}(t,k)$ is a smooth function of t. )

 Thus, it follows that
\begin{equation}
\label{eq3}
\nu(P(1)) = \nu(P(0)) = \nu(P).
\end{equation}
Further, since $S(1)$ can be regarded as the union of the planes  $k_{y}= \pi$ and
$k_{z}=0$, 
\begin{equation}
\nu(P(1))= \tilde{\nu}_{2} + \nu_{3}.
\label{eq:1}
\end{equation} 

Thus, from Eqs.~(\ref{eq2}),(\ref{eq3}) and (\ref{eq:1}), we conclude that 
\[
\nu_{2}-\tilde{\nu}_{2}= \nu_{3}- \tilde{\nu}_{3}.
\]  

The plane $k_{z}=k_{y}$ can also be obtained as a deformation of the
surface in Fig.~\ref{fig1}(b). Thus the $Z_{2}$ invariant of this
plane is obtained as \( \tilde{\nu}_{2} + \nu_{3} \). Similarly, by deforming
surfaces $S'$ which we define as the union of the planes $k_{x}=0 $
and $k_{y}=\pi$ and $S''$ which we define as the union of the planes
$k_{x}=0$ and $k_{z}=\pi$ it can easily be shown that
\begin{equation}
\nu_{1}- \tilde{\nu}_{1}=\nu_{2}-\tilde{\nu}_{2}= \nu_{3}- \tilde{\nu}_{3}.
\end{equation}

Further, the $Z_{2}$ invariant of any other plane which maps onto
itself under TRS can also be obtained from the values of
$\nu_{1},\nu_{2},\nu_{3}$ and $\nu_{1}-\tilde{\nu}_{1}$. We can thus
characterize any topological insulator with TRS in three dimensions
with four invariants, which may be chosen to be
\(\nu_{1},\nu_{2},\nu_{3}\) and \(|\nu_{1}-\tilde{\nu}_{1}| \).

 The arguments in Ref.~\onlinecite{roy_topological_2009} may be summarized as
follows.  The spectral projection operator in 3d when restricted to
the planes $k_{i}=0,k_{i}=\pi$ for $k_{i}\in \{k_{x},k_{y},k_{z}\}$ can be written as a sum 
$P = P_{1}+ P_{2}$, where $P_{1}, P_{2}$ map onto each other under TRS.

Consider the set of momentum space slices, $k_{z}= c$, where $ 0\le c
\le \pi$.  Let $P_1(c), P_2(c)$ be the restriction of the operators
$P_1$ and $P_2$ to the plane $k_z=c$. For an arbitrary value of $c$,
these operators do not map onto each other under TRS. The $Z_2$
invariant is therefore not well defined in general. When the $Z_{2}$
invariants for the planes $k_{z}=0$, $k_{z}= \pi$ are different, the
operators, $P_{1}(0)$ and $P_1(\pi)$ have different Chern numbers. A
continuous deformation of a two dimensional projection operator to one
which has a different Chern number is not possible. Changes in Chern
number may be regarded as occurring at singular diabolical points or
monopoles at which the projection operator is not well defined.  In
general, one may define 2d surfaces enclosing points at which the
projection operators $P_1,P_2$ are not well defined and associate a
charge with these surfaces. If the difference in Chern number of
$P_{1}$ on the slice $k_{z}=0$ to the Chern number of $P_{1}$ on the
slice $k_{z}=\pi$ is an odd integer, this implies the existence of a
net odd monopole charge between these two planes and ensures that the
$Z_{2}$ invariants of the planes $k_{z}=0$ and $k_{z}=\pi$ are
different. By a careful counting of monopole charges and by using TRS,
one can then show that in this case, the $Z_{2}$ invariants of the
$k_{x}=0$ and $k_{x}=\pi$ planes also differ from each other and the
same is true for the $k_{y}=0$ and $k_{y}=\pi$ planes.

 This fourth $Z_{2}$ invariant, $\nu_{1}- \tilde{\nu}_{1}$ is an
intrinsically three dimensional characteristic of the insulator. 
Insulators whose fourth $Z_{2}$ invariant is zero and one have been
christened ``weak'' and ``strong'' topological insulators respectively. 
The four invariants found here agree with the counting in Refs.~\onlinecite{
moore2006tit,fu2007tit}.

 In summary, we have analyzed the Chern number formula for the
 $Z_{2}$ invariant in two dimensions. The trivial and non-trivial
 topological classes may be thought of as equivalence classes of
 Hamiltonians which contain respectively members with
 even and an odd quantized spin Hall conductance. A simple counting
 argument for the number of invariants in 3d was provided using 
 deformations of planes which map onto themselves under time reversal.

 The author is grateful to John Chalker, Dmitry Kovrizhin and Steven
 Simon for useful discussions and comments on previous versions of
 this manuscript and acknowledges support from EPSRC grant
 EP/D050952/1.

\end{document}